\documentclass[a4paper,aps,preprintnumbers,twocolumn,amsmath,amssymb,pra]{revtex4}
\usepackage{bbm}
\usepackage{mathrsfs}
\usepackage{amsmath}
\usepackage[dvips]{graphicx}
\usepackage{epsfig}
\usepackage[dvips]{graphics,graphicx}

\newcommand{\be}{\begin{equation}}
\newcommand{\ee}{\end{equation}}
\newcommand{\bea}{\begin{eqnarray}}
\newcommand{\eea}{\end{eqnarray}}
\newcommand{\bes}{\begin{split}}
\newcommand{\ees}{\end{split}}

\newcommand{\Vk}{\mathbf{k}}

\newcommand{\sech}{\mbox{sech}}

\usepackage[dvips]{graphics,graphicx}

\begin{document}
\title{Damping in 2D and 3D dilute  Bose gases}
\author{ Ming-Chiang Chung$^1$ and Aranya B. Bhattacherjee$^2$}
\affiliation{$^1$Institute of Physics, Academia Sinica, Taipei 11529, Taiwan}
\affiliation{$^2$Department of Physics, ARSD College, University of Delhi (South Campus), New Delhi-110021, India}

\begin{abstract}
 Damping in  2D and 3D dilute gases is investigated using both the hydrodynamical approach and the Hartree-Fock-Bogoliubov (HFB) approximation . We found that the both methods are good for the Beliaev damping at zero temperature and Landau damping at very low temperature, however, at high temperature, the hydrodynamical approach overestimates the Landau damping and the HFB gives a better approximation. This result shows  that the comparison of the theoretical calculation using the hydrodynamical approach and the experimental data for high temperature done by Vincent Liu (PRL {\bf21} 4056 (1997)) is not proper. For two-dimensional systems, we show that the Beliaev damping rate is proportional to $k^3$ and the Landau damping rate is proportional to $ T^2$ for low temperature and to $T$ for high temperature. We also show that in two dimensions the hydrodynamical approach gives the same result for zero temperature and for low temperature as HFB, but overestimates the Landau damping for high temperature.
\end{abstract}

\date{\today}
\maketitle

\section{Introduction}
The experimental realization of Bose-Einstein condensation (BEC)  in magnetically trapped akali atoms \cite{Anderson,Davis,Bradley} provides a good tool to study the properties of 3D dilute Bose gases. Furthermore, with the anisotropic of new traps \cite{Goerlitz, Smith}, one can further confine condensate atoms in a quasi-two-dimensional regime \cite{Smith,Stock}.  Most theoretical work (for 3D and 2D systems, see Review \cite{Review3d,Review2d}, respectively) has focused on the dynamics of condensates and the zero-temperature behavior, which can be obtained  by solving the non-linear Gross-Pitaevskii (GP) equation.  However, the finite-temperature behavior has still remained difficult to study, where experiments have shown damping of the condensates modes in the presence of a significant noncondensate component \cite{Jin, Miesner}. 
Damping mechanism associated with collective excitations of Bose condensed atoms interacting with a non-condensed, thermal component is not well understood and still represents a challenging problem in theoretical physics. The damping of collective modes can have various origins. There are two distinct contributions to the total decay rate $\gamma=\gamma_{B}+\gamma_{L}$: One arises at $T=0$ from the process of decay of a quantum of excitation into two or more excitations with lower energy. This mechanism was first studied by Baliaev \cite{Beliaev}in 3D uniform Bose superfluids and is known as Balieav damping $\gamma_{B}$. At finite temperature, a different mechanism of damping (known as Landau damping, $\gamma_{L}$) comes from the process of one quantum of excitation decays due to coupling with transitions associated with other elementary excitations and occurring
  at the same frequency. Landau damping is not associated with thermalization process and can be well described in
  the framework of mean field theory \cite{Pitaevskii,Giorgini,PitaGuill}. The subject of Landau damping in dilute BECs has been explored by several authors. Landau damping in an uniform Bose gas at low temteratures was first investigated by Popov \cite{Popov} Hohenberg and Martin \cite{Hohenberg}, while at high temperatures was first investigated by Sz`{\'e}pfalusy and Kondor \cite{Szepfalusy}. The relevance of Landau damping to explain experimental data of trapped Bose gas was proposed by Liu and Schieve \cite{Schieve} and developed by Liu \cite{Liu} using the Popov's hydrodynamical approach \cite{Popov}. On the othe hand,  Pitaevskii and Stringari \cite{Pitaevskii} investigated Landau damping in a weakly interacting uniform as well as non-uniform Bose gas by means of semi-classical theory.
 They showed that for the uniform Bose gas, it reproduces known results for both the low  temperature asymptotic behaviour of the phonon coupling.
However, for the high temperature, Liu showed higher Landau damping rate than those obtained by  Sz{\'e}pfalusy and Kondor, while the hign-temperature behavior could be reproduced by Pitaevskii and Stringari.

However all investigations of the damping rate has been done for a 3D Bose gas. 2D Bose gases are interesting as their low temperature physics is governed by strong long-range fluctuations. These fluctuations inhibit the formation of true long-range order, which is a key concept of phase transition theory in 3D. Thus a 2D uniform interacting Bose gas does not undergo Bose-Einstein condensation at finite temperatures. However, this system turns superfluid below the BKT (Berezinski, Kosterlitz and Thouless) temperature $T_{KT}$ \cite{Berezinski,Kosterlitz}. The experiment indication of the BKT transition in weakly interacting Bose system has even been shown in Ref. \cite{Stock}. Damping in a 2D Bose gas is an open question which was recently addressed by the authors for a uniform Bose gas \cite{Ming} using the hydrodynamical theory of Popov \cite{Popov}. In this work, we show that the hydrodynamical approach actually overestimates the damping rate at high temperatures, both for 3D and 2D systems and we calculate the Balieav and Landau damping rates for a 2D uniform Bose gas using the semi-classical
  Hartree-Fock-Bogoliubov (HFB) approach. In the limit of low temperatures, the results of this approach is in good agreement with that found from hydrodynamical approach. Contrary to earlier work \cite{Liu}, we show that for the 3D case in the high temperature limit, the hydrodynamical approach cannot be used to explain the experimental data.

This paper is organized as follows. In Sec. \ref{sec2} we discuss the relation between the atom-atom interaction and the scattering length for 2D and 3D dilute gases. In Sec. \ref{sec3} we first introduce the hydrodynamical approach developed by Popov\cite{Popov}, and then calculate the Beliaev damping and Landau damping for 3D and 2D gases. The mistake using this approach for high temperature is also discussed. In Sec. \ref{sec4} the HFB approximation is developed to calculate 3D and 2D Beliaev and Landau damping.

\section{Atom-atom Interaction and scattering Length} \label{sec2}
The standard Hamiltonian of an interacting Bose gas is
\be
 \begin{split}
   H  = &  \int  {\mbox{d}}^d{\mathbf r} \frac{1}{2} \nabla \psi^{\dagger}({\mathbf r}) \nabla \psi^{\dagger}({\mathbf r})  + V_{\mbox{ext}}({\mathbf r}) \psi^{\dagger}({\mathbf r}) \psi({\mathbf r}) \\
     + &   \frac{1}{2} \int   {\mbox{d}}^d{\mathbf r}
     {\mbox{d}}^d{\mathbf r}'   \psi^{\dagger}({\mathbf r})  \psi^{\dagger}({\mathbf r}') U({\mathbf r} - {\mathbf r}')  \psi({\mathbf r})  \psi({\mathbf r}'),
  \end{split}
\ee
where $U$ is the atom-atom interaction and $V_{\mbox{ext}}$ is the external potential. For uniform Bose gas, $V_{\mbox{ext}} = 0$.
The true interaction between atoms is very complicated where one has to consider the fine structure of atoms. However, the scattering process can offer a effective potential to simplify the ineraction. In order to do that, one has to introduce Green functions for bosonic systems with condensate.
The difficulty of doing so  arises from the fact that the terms containing the odd number of annihilation operators do not vanish for a Bose gas  after averaging the ground state due to the existence of condensate, which unfortunately destroys the hope to apply the normal technique of Feynman diagrams to the system. This difficulty was successfully resolved by Beliaev \cite{Beliaev, Beliaev2}. He separated the operators with zero momentum, which semi-classically can be regarded as a c-number, and the other operators with nonzero momenta. In this way, the Feynman diagrams can be used for the Bose gas.

Beliaev considered a three-dimensional system with short-range, central interaction potential with radius $1/a_3$ and then calculated the renormalized atom-atom interaction in the presence of the condensate  between two particles with non-zero momenta, which one should sum over all ladder diagrams. In this way, one can obtain the renormalized interaction in terms of the s-wave scattering amplitude according to the elementary scattering theory \cite{Fetter, Review3d, Leggett}. Therefore the effective potential can be written as
\be
  U({\mathbf r}) = \frac{4\pi a_3}{m} \delta({\mathbf r}),
\ee
with the  atom mass $m$,
   and the momentum dependence of the scattering amplitude can be ignored in the low temperature limit. In the rest of the paper we define the atom-atom interaction strength $g_3$ as
\be
  g_3 =  \frac{4\pi a_3}{m}.
\ee

For two dimensions, Schick followed the methods developed by Beliaev and examined a two-dimensional systems of  hard-core bosons with a diameter $a_2$ at low density and zero temperature.  Unlike the three-dimensional systems, where ladder diagrams are independent of the dimensionless parameter $na_3^3$, hence, it is natural to take it as the small perturbation terms to expand the quantities.  For two-dimensional systems, contributions from the ladder diagrams depends logarithmically on $n a_2^2$, the dimensionless parameter for 2D systems, but not directly on $na_2^2$ itself.  In particular, the renormalized interaction is proportional to   $1/\ln{1/na_2^2}$:
\be
 g_2 = \frac{4 \pi}{m\ln{(1/na_2^2)}}.
\ee
Schick concluded that the $1/\ln{1/na_2^2}$ plays a role of a small parameter in the  two-dimensional dilute systems, and other quantities, like damping rate in this paper,  can be expanded in terms of it.

\section{Hydrodynamical  approach} \label{sec3}
In the low temperature and low energy limit, Popov \cite{Popov}  developed a hydrodynamical  approach to find an effective Hamiltonian  for a nonideal Bose gas. In order to do that, one has to separate the order parameter over rapidly and slowly oscillating field, and the hydrodynamical Hamiltonian can be obtained by integrating the functional over rapidly oscillating field. The theory describes then the hydrodynamical Hamiltonian in terms of two slowly varying fields : phase $\phi(x)$ and density fluctuation $\pi(x) = n(x) - n_0 $ with $n_0$: the density of the ground state.
Here the four-dimension  Euclidean space $x=(\mathbf{x}, \tau)$ is used with the imaginary time $\tau$.  The hydrodynamical  action for a $d-$dimensional nonideal Bose gas, according to Popov, can be written in the form (notice that $\hbar = 1$ throughout the paper)
\be
\begin{split}
S[\phi,\pi] & = \int_0^\beta {\mbox{d}} \tau \int  {\mbox{d}}^d x
\left\{ i{\partial^2 p \over \partial \mu \partial n} \pi
\partial_\tau \phi  -{1\over 2m} {\partial p\over \partial \mu}
 (\nabla\phi)^2  \right. \\
  & \left. - {1\over 2}{\partial^2 p\over \partial
\mu^2} (\partial_\tau\phi)^2
+ {1\over 2}{\partial^2 p\over
\partial n^2} \pi^2 - {(\nabla\pi)^2
\over 8mn_0} - {\pi (\nabla\phi)^2 \over 2m}
\right\}
\label{eq:Seff}
\end{split}
\ee
with the atom mass $m$, the pressure of a homogeneous   system $p$,  the chemical potential $\mu$ and the atom density $n$. The fields $\phi$ and $\pi$ are periodic in imaginary time $\tau$ with the period $\beta = 1/(k_B T)$.  For very low temperature, as long as the non-condensate part  can be neglected compared to the condensate, the pressure $p(\mu, n )$ at zero temperature can be a good approximation. Therefore we can use the expression of a weakly interacting dilute gas as
     $p = \mu n - \frac{g_d}{2} n_0^2,$
where $g_d$ is the atom-atom interaction related to the  scattering length.   It follows that
\be
  {\partial^2 p \over \partial \mu \partial n} = 1; \;\;\; \frac{\partial p}{\partial \mu} = n  \simeq n_0; \;\;\; \frac{\partial^2 p}{ \partial \mu^2} = 0;
\;\;\; {\partial^2 p \over \partial n^2} = -g_d,
\ee
and the action (\ref{eq:Seff}) takes the form

\be \label{eq:Seff2}
   \int {\mbox{d}} \tau {\mbox{d}}^d x \left(i \pi \partial_{\tau} \phi -  \frac{(n_0 + \pi)}{2m}  (\nabla \phi)^2 - \frac{g_d}{2} \pi^2 - \frac{(\nabla \pi)^2}{8mn_0} \right).
\ee
 The action contains all quadratic functions except the term $ 1/{2m}  \pi (\nabla \phi)^2$, considered as an interacting potential.  The Hamiltonian can be derived from the effective action  (\ref{eq:Seff2}) as
\be \label{eq:SeffLT}
      \int  {\mbox{d}}^d x \left( \frac{m}{2} n \mathbf{v} ^2 + \frac{g_d}{2} (n-n_0)^2 + \frac{(\nabla n)^2}{8mg_0}  \right),
\ee
where the  field of velocities is defined as  $ \mathbf{v} = 1/m \nabla \phi$. This Hamiltonian is consistent with one particular realization of the Landau hydrodynamical Hamiltonian.

\begin{figure}
\center
\includegraphics[width=8.0cm]{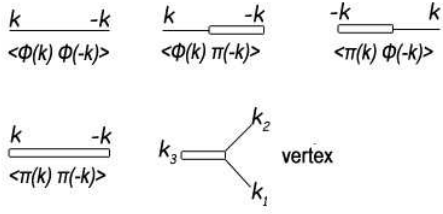}
\caption{Green's functions and vertex} \label{fig1}
\end{figure}

Fourier transforming the fields $\phi$ and $\pi$, the effective action (\ref{eq:SeffLT}) can be written in the form
\be  \label{eq:SeffMS}
\begin{split}
    -{1\over 2 } \sum_{\nu} & \int {\mbox{d}}^d k  \frac{n_0}{m} {\mathbf{k}}^2 \phi(k) \phi(-k) + 2 \omega_{\nu} \phi(k) \pi(-k) \\ & + (g_d + \frac{\mathbf{k}^2}{4mn_0}) \pi(k) \pi(-k ) \\ & -
     \frac{1}{\sqrt{\beta V}} \sum_{k_1 + k_2 + k_3 = 0} \frac{\mathbf{k}_1 \cdot \mathbf{k}_2 }{2m} \phi(k_1) \phi(k_2) \pi(k_3),
\end{split}
\ee
where  $k$ is the vector $(\mathbf{k}, i\omega_{\nu})$  and the Matsubara frequencies $\omega_{\nu} = 2 \pi \nu / \beta$ with integers $\nu$.
From the action of the Fourier transformation (\ref{eq:SeffMS}) one can extract the important information needed for the perturbation calculations using diagrammatic technique. First of all, the free Green's functions is defined as follows
\be
 G_0(k ) = \left( \begin{array}{cc}  \langle \phi(k) \phi(-k) \rangle_0  &  \langle \phi(k) \pi(-k) \rangle_0  \\  \langle \pi(k) \phi(-k) \rangle_0   &\langle \pi(k) \pi(-k) \rangle_0, \end{array}\right)
\ee
where $\langle \cdots \rangle_0$ denotes the expectation value of fields calculated only with the quadratic action.   From the action  (\ref{eq:SeffMS}), the inverse of the free Green's function can be found as
\be\label{eq:BG}
     G^{-1}_0(k) =  \left( \begin{array}{cc}   \frac{n_0}{m} \mathbf{k}^2 &  \omega_{\nu}  \\  -\omega_{\nu}%
  & g_d + \frac{\mathbf{k}^2}{4mn_0}
\end{array} \right).
\ee
Therefore,
\be
 G_0(k) = \left(
\begin{array}{cc}
{g_d + \mathbf{k}^2/(4mn_0) \over \omega_\nu^2 +\epsilon^2(\mathbf{k})} &
{-\omega_\nu \over \omega_\nu^2 +\epsilon^2(\mathbf{k})} \\
{\omega_\nu \over \omega_\nu^2 +\epsilon^2(\mathbf{k})} &
{(n_0/m) \mathbf{k}^2 \over \omega_\nu^2 +\epsilon^2(\mathbf{k})}
\end{array} \right),
\label{eq:G_0}
\ee
where
\be \label{BogoliubovSpectrum}
 \epsilon(\mathbf{k}) = \sqrt{(\frac{\mathbf{k}^2}{2m})^2 + c^2 \mathbf{k}^2} \ee
 with $c \equiv \sqrt{g_d n_0/m}$.  We represent the relation between the free Green's functions and  the Feyman diagrams in Fig.\ref{fig1}.   The cubic term of the action  (\ref{eq:SeffMS}) is known as phonon-phonon  interaction in the low temperature region, giving rise to a vertex of $\delta^d({\mathbf{k}_1 +\mathbf{k}_2 +\mathbf{k}_3})
\delta_{\nu_1+\nu_2+\nu_3, 0} {({\mathbf{k}_1 \cdot \mathbf{k}_2}) \over m}$ represented by the last diagram of Fig. \ref{fig1}.

\begin{figure}
\center
\includegraphics[width=8.0cm]{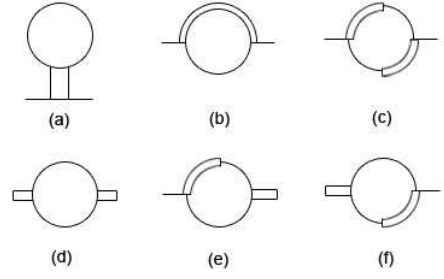}
\caption{One-loop diagrams for the self energy.} \label{fig2}
\end{figure}

The exact Green's function has to involve the phonon-phonon interaction, given by the Dyson equation $G(k)  = G_0 (k) + G_0(k) \Sigma(k) G(k)$, where $\Sigma(k)$ represents the self-energy matrix. The low-frequency spectrum of collective modes  can be obtained by the poles of the exact Green's function as
\be \label{eq:PolesG}
  \mbox{det} G^{-1}(k)  = \mbox{det} [G_0^{-1}(k))-\Sigma(k)] = 0,
\ee
through  the analytical continuation $i\omega_{\nu} = \omega + i \eta $ ($\eta = 0^+$) after the Matsubara frequency sum. The complex frequency $\omega = E  - i \gamma(\mathbf{k})$ represents the energy spectrum $E$ and the damping rate $\gamma$.  Neglecting the matrix $\Sigma$, the zero order approximation for the Eqs. (\ref{eq:PolesG}) gives the square of the  Bogoliubov energy spectrum:
\be \label{eq:BogoE}
   E^2 =  (\frac{\mathbf{k}^2}{2m})^2 + c^2 \mathbf{k}^2 = \epsilon(\mathbf{k})^2
\ee
When the phonon-phonon interaction is considered, the imaginary part appears in the spectrum. Fig. \ref{fig2}  shows the one-loop diagrams for the self energy $\Sigma$. The contribution to the imaginary part of the spectrum is given by the last five Figs. \ref{fig2}(b) - \ref{fig2}(f). The damping rate can be obtained from Eqs. (\ref{eq:BG}) and (\ref{eq:PolesG}) \cite{Liu} as
\begin{equation}
\begin{split}
\gamma(\Vk)=&  {1\over 2 E}
\left[(g_d+{\Vk^2\over 4mn_0}) \mbox{Im}\Sigma_{\phi\phi}(\Vk,\omega +i\eta) \right.\\ & \left. +
{n_0\Vk^2 \over m} \mbox{Im}\Sigma_{\pi\pi}(\Vk,\omega+i\eta)
\right ] \\  & - \mbox{Re}\Sigma_{\phi\pi}(\Vk,\omega+i\eta).
\end{split}
\label{eq:gamma}
\end{equation}
Replacing (\ref{eq:BogoE}) into Eqs. (\ref{eq:gamma}) and calculating the diagrams, we obtain
$
  \gamma(\Vk) = \gamma_B(\Vk) + \gamma_L(\Vk),
$
where $\gamma_B$ is the Beliaev damping as
 \be\label{eq:BeliaevD}
 \begin{split}
& \gamma_B(\Vk) =  {1\over  2^{d+2}\pi^{d-1}}\int \mbox{d}^d k^\prime
 \delta(\epsilon(\Vk)-\epsilon(\Vk^\prime) -\epsilon(\Vk-\Vk^\prime)) \\
& [f^0(\epsilon(\Vk^\prime)) -f^0(-\epsilon(\Vk-\Vk^\prime))]
 \left\{
{(\Vk-\Vk^{\prime})^2 (\Vk\cdot\Vk^\prime)^2 \epsilon(\Vk^\prime) \epsilon(\Vk)
\over 2mn_0 \Vk^2{\Vk^\prime}^2  \epsilon(\Vk-\Vk^\prime) }\right. \\
 &    +{(\Vk\cdot\Vk^\prime) (\Vk\cdot(\Vk-\Vk^\prime))\epsilon(\Vk)
\over 2mn_0 \Vk^2}
\\ &+{\Vk^2 (\Vk^\prime\cdot(\Vk-\Vk^\prime))^2
\epsilon(\Vk^\prime)\epsilon(\Vk-\Vk^\prime) \over 4 mn_0
{\Vk^\prime}^2 (\Vk-\Vk^\prime)^2\epsilon(\Vk)} \\
& \left. +
{(\Vk^\prime\cdot(\Vk-\Vk^\prime))(\Vk\cdot\Vk^\prime)\epsilon(\Vk^\prime)
\over mn_0{\Vk^\prime}^2} \right\}
\end{split}
\ee
and $\gamma_L$ is known as Landau damping as
\be \label{eq:LandauD}
\begin{split}
& \gamma_L(\Vk) = {1\over 2^{d+2}\pi^{d-1}}\int {\mbox{d}}^d k^\prime
\delta(\epsilon(\Vk)+\epsilon(\Vk^\prime) -\epsilon(\Vk+\Vk^\prime)) \\
& [f^0(\epsilon(\Vk^\prime)) -f^0(\epsilon(\Vk+\Vk^\prime)) ]
  \\ & \left\{{\epsilon(\Vk) \over 2mn_0} \left[
{{\Vk^\prime}^2 (\Vk\cdot(\Vk+\Vk^\prime))^2 \epsilon(\Vk+\Vk^\prime)
\over  \Vk^2(\Vk+\Vk^\prime)^2\epsilon(\Vk^\prime)} \right. \right. \\
&\left.  + {(\Vk+\Vk^\prime)^2 (\Vk\cdot\Vk^\prime)^2 \epsilon(\Vk^\prime)
\over  \Vk^2{\Vk^\prime}^2\epsilon(\Vk+\Vk^\prime)} \right]
 +{(\Vk\cdot\Vk^\prime) (\Vk\cdot(\Vk+\Vk^\prime))\epsilon(\Vk)
\over mn_0 \Vk^2}  \\
& +{\Vk^2 (\Vk^\prime\cdot(\Vk+\Vk^\prime))^2
\epsilon(\Vk^\prime)\epsilon(\Vk+\Vk^\prime) \over 2 mn_0
{\Vk^\prime}^2 (\Vk+\Vk^\prime)^2\epsilon(\Vk)}
   + {(\Vk^\prime\cdot(\Vk+\Vk^\prime))\over mn_0} \\
 & \left. \times \left[
  {(\Vk\cdot\Vk^\prime)\epsilon(\Vk^\prime) \over {\Vk^\prime}^2} +
  {(\Vk\cdot(\Vk+\Vk^\prime))\epsilon(\Vk+\Vk^\prime) \over
(\Vk+\Vk^\prime)^2 } \right] \right\},
\end{split}
\ee
where the bosonic distribution function $f^0(\epsilon) = 1/[\exp(\beta \epsilon) -1]$.

\subsection{Quantum regime $ck \gg k_BT$}
At $T = 0$, the Landau damping disappears and the Beliaev damping contributes to the damping rate. In the Beliaev damping mechanism the momenta of the three excitations are comparable, $|\Vk| \simeq |\Vk^{\prime}| \simeq |\Vk - \Vk^{\prime}|$. Then the Eqs. (\ref{eq:BeliaevD}) yields
\be \label{eq:BD}
  \begin{split}
   \gamma_B(\Vk) = & {9 c \over 2^{d+4} \pi^{d-1} mn_0} \int \mbox{d}^dk^{\prime} |\Vk||\Vk^{\prime}|  |\Vk - \Vk^{\prime}| \\ &  \delta(\epsilon(\Vk)-\epsilon(\Vk^\prime) -\epsilon(\Vk-\Vk^\prime)).
\end{split}
\ee
 In three dimensions the  damping rate for small $k$ ($k \ll mc$) is
\be \label{3DBD}
   \gamma_B^{d=3}(\Vk) = {3 k^5 \over 640 \pi mn_0 },
\ee
known as Beliaev's result\cite{Beliaev}.

For two-dimensional systems, the Eqs. (\ref{eq:BD}) can be written as
\be
  \gamma_{B}^{d=2}(\Vk) = 2\left[{9 \over 64 \pi mn_0} \int \mbox{d}k^{\prime} {|\Vk^{\prime}| (|\Vk - \Vk^{\prime}|)^2 \over \sin\theta}\right],
\ee
where $\theta$ is the angle between $\Vk$ and $\Vk^{\prime}$. The factor two in front of the bracket comes from the fact that there are two angles corresponding to  the energy conservation for the Beliaev damping ($\epsilon(\Vk)-\epsilon(\Vk^\prime) = \epsilon(\Vk-\Vk^\prime)$): $\sin\theta \simeq \pm {\sqrt{3}  |\Vk - \Vk^{\prime}| \over 2mc }$ \cite{Semenov}, and the Beliaev damping rate for a $2$-D Bose gas has the form
\be \label{2DBD}
  \gamma_{B}^{d=2} = {\sqrt{3} c \over 32 \pi n_0}  k^{3}.
\ee
This result corrects the wrong result previously given by Chung and Bhattacherjee \cite{Ming} and the factor two will appear naturally in the two-dimensional Landau damping. 

\subsection{Thermal Regime $ck \gg k_BT$}
For finite temperature and small momenta such that $cq \ll k_B T$ and $cq \ll
 n_0 g_d $, the Beliaev damping is much smaller than the Landau damping.
 In three dimensions, the damping rate to the lowest order in $k$ can be obtained from Eq. (\ref{eq:LandauD}) as
\be
   \frac{\gamma_L^{d=3}(k)}{\epsilon(k)} =  \frac{k_0^5}{16\pi m n_0 k_B T} {\cal I}_3(\tau),
\ee
 where
\be \label{3DDampT}
  \begin{split}
   {\cal I}_3(\tau) = & \frac{1}{4}\int_0^{\infty}
            dz z^2 \sech^{2}(\frac{z}{2\tau})  \\
       &  [\frac{1}{2} + \frac{3}{2(z^2+1)} + \frac{2z^2}{(z^2+1)^2} - \frac{2}{(z^2+1)^2}],
   \end{split}
\ee
 with $k_0 = \sqrt{mn_0g_3}$ and $\tau \equiv k_BT /n_0 g_3 $ . This result was first obtained by V. Liu \cite{Liu}.  For $k_B T  \ll mc^2$, Eq.{\ref{3DDampT}} is reduced to the Hohenberg and Martin's  result \cite{Hohenberg}
\be \label{Hohenberg}
  \gamma_L^{d=3}(k) = \frac{3\pi^3 k (k_BT)^4}{40 mn_0c^4}.
\ee
This low temperature limit gives  the same result as that using the HFB approach, which will be introduced in the next section. For high temperature $k_B T \gg mc^2$, ${\cal I}(\tau) \sim 38.735 \tau$, and the damping rate is approximated by
\be
  \frac{\gamma_L^{d=3}(k)}{\epsilon(k)} \simeq 9.648 \frac{k_B T a_3}{c}
\ee
with the three-dimensional scattering length $a_3 = m g_3/4 \pi$. Unfortunately this result is different than that investigated by Sz{\'e}pfalusy and Kondor
\cite{Szepfalusy}, which reads
\be \label{Kondor}
  \frac{\gamma_L^{d=3}(k)}{\epsilon(k)} \simeq \frac{3\pi}{8} \frac{k_B T a_3}{c}.
\ee
Therefore the hydrodynamical approach is no longer correct for the high temperature.
 The reason is that in the hydrodynamical Hamiltonian only the slow oscillating fields are considered by integrating out the fast oscillating fields. For high temperature the fast oscillating fields should also be considered to reduce the damping rate.  We can conclude that the hydrodynamical approach is very good for low temperature, however, for high temperature, other method should be introduced. We will discuss that in the next section.

\begin{figure}
\center
\includegraphics[width=8.0cm]{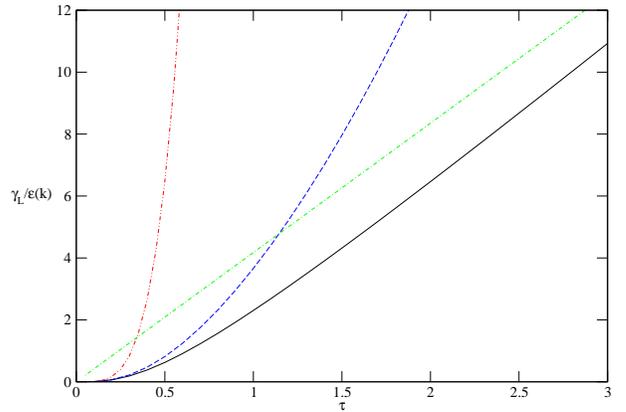}
\caption{Landau damping rate per unit energy versus $\tau$ in three dimensions. The unit of $\gamma_L/\epsilon({\mathbf k})$ is $\sqrt{a_3^3n_o}$. Solid black line represents the result obtained by HFB method, dashed blue line by the hydrodynamical approach. The high-temperature asymptotic behavior and low-temperature limit are also shown by dashed-dot green line and dashed-dot-dot red line, respectively.} \label{fig3}
\end{figure}

In Fig.~\ref{fig3} the three-dimensional Landau damping  per unit energy using the hydrodynamical approach (dashed line) and HFB (solid line) is plotted as a function of $\tau$. Also shown are  the asymptotic behavior at high temperature (the dashed-dot line)  and the low temperature limit (the dashed-dot-dot line).   We can see that the hydrodynamical approach gives very good agreement with the low temperature limit  for $\tau \leq 0.5$,  however, it goes too large at high temperature and it does not approach the asymptotic value given by Szepfalusy and Kondor. Therefore the conclusion made by Liu in Ref. \cite{Liu} that the results of the hydrodynamical approach can fit the experimental data is not proper.

In two dimensions, the damping rate reads
\be
    \frac{\gamma_L^{d=2}(k)}{\epsilon(k)} =  \frac{\sqrt{2} k_0^4}{16mn_0k_BT} {\cal I}_2(\tau),
\ee
where
\be
   \begin{split}
      {\cal I}_2(\tau) = &  \frac{1}{4}\int_0^{\infty}
            dz z^2 \frac{\sech^{2}(\frac{z}{2\tau})}{\sqrt{\sqrt{z^2+1}-1-\frac{z^2}{2(z^2-1)}}} \\
     &  [\frac{1}{2} + \frac{3}{2(z^2+1)} + \frac{2z^2}{(z^2+1)^2} - \frac{2}{(z^2+1)^2}].
   \end{split}
\ee
In the low temperature limit: $k_B T  \ll mc^2$, ${\cal I}_2(\tau) \rightarrow \sqrt{6} \pi^2 \tau^2$, therefore the damping coefficient is given by
\be \label{2DLDLT}
 \frac{\gamma_L^{d=2}}{\epsilon(k)} = \frac{\sqrt{3} \pi}{8} \frac{(k_BT)^2}{n_0c^2}.
\ee
 In this low temperature regime, the damping rate is proportional to $T^2$. As far as we know, this quadratic dependence of the temperature for the damping rate in the low temperature is found for the first time in this paper.

For the high temperature, as the three-dimensional case, the hydrodynamic Hamiltonian  overestimates the damping. In the next section, we will use the  Hartree-Fock-Bogoliubov approximation to obtain the two-dimensional  damping at high temperature.

\begin{figure}
\center
\includegraphics[width=8.0cm]{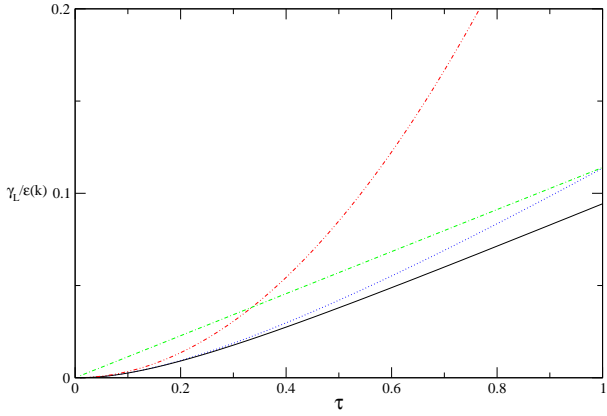}
\caption{Landau damping rate per unit energy in two dimensions in the unit of $8\pi/|\ln na_2^2 |$. The black solid line represents the result from HFB and the blue dashed line from the hydrodynamical approach. The high-temperature asymptotic behavior and the low-temperature limit are also reported as the green dashed-dot line and the red dashed-dot-dot line, respectively.} \label{fig4}
\end{figure}

Fig. (\ref{fig4}) shows the two-dimensional Landau damping rate per unit energy using both hydrodynamical approach (dashed line)  and HFB method (solid line). In this figure the low temperature limit (dashed-dot-dot line) and the asymptotic value  at high temperature (dashed-dot line)  are also shown.  The hydrodynamical result is in agreement with the low temperature limit for $\tau < 0.2$, however, it will not approach the asymptotic value for high temperature similar to  the three-dimensional case.

%
%

\section{ Hartree-Fock-Bogoliubov Approach} \label{sec4}
In this section we represent a semi-classical method: Hartree-Folk-Bogolubov (HFB) . We will see that in the low-temperature regime this approach is in a good agreement  with the hydrodynamical approach, while for the high temperature, on the contrary to the hydrodynamic approach, HFB gives a better approximation to the decay rate.

We start with the method by Giorgini \cite{Giorgini}.  The grand-canonical Hamiltonian of a system with a nonuniform external field $V_{\mbox{ext}}({\mathbf r})$ reads
\be \label{KOrigin}
  \begin{split}
   K = & H - \mu N =  \int {\mbox{d}}^d{\mathbf r} \psi^{\dagger}({\mathbf r},t) H_0 \psi({\mathbf r},t)  \\
      & + \frac{g_d}{2} \int d{\mathbf r} \psi^{\dagger}({\mathbf r},t) \psi^{\dagger}({\mathbf r},t) \psi({\mathbf r},t) \psi({\mathbf r},t),
  \end{split}
\ee
where
\be
 H_0 = - \frac{\nabla^2}{2m} + V_{\mbox{ext}}({\mathbf r}) - \mu
\ee
and $\psi^{\dagger}({\mathbf r},t)$ and $\psi({\mathbf r},t)$ are the creation and annihilation field operators.
Since the system is in the regime where the condensate exists, we  define a time-dependent condensate wave function $\Phi({\mathbf r},t)$ \cite{Hohenberg}
\be \label{TDCond}
  \Phi({\mathbf r},t) = \langle \psi({\mathbf r},t) \rangle
\ee
 with the average $\langle \cdots \rangle$ using the grand-canonical Hamiltonian
(\ref{KOrigin}). We have to notice that the Eq.~(\ref{TDCond}) can always be used if the condensate exists, however, for a homogeneous two-dimensional system,
i.e.   $ V_{\mbox{ext}}({\mathbf r}) = 0$, the condensate does not exist at finite temperature. In this case the long-range order disappears, it remains the quasi-long-range order for  a two-dimensional homogeneous Bose gas. That means, though a macroscopic occupation number of a single state does not exist, there exists a small value of $k_c$ in the momentum space where a macroscopic occupation number of the states $k < k_c$ still forms a quasi-condensate.
  Therefore the bracket in Eq.~(\ref{TDCond}) should count all the states in the quasi-condensate. We can see that $\Phi({\mathbf r},t)$ allows us to describe the oscillating condensate away from the equilibrium. To avoid the confusion for the notation, we define here the stationary value of the condensate in its equilibrium as $\Phi_0({\mathbf r})$
\be \label{Decomp}
  \Phi_0({\mathbf r}) = \langle \psi({\mathbf r}) \rangle_0,
\ee
where $\langle \cdots \rangle_0$ denotes the time-independent  average of the condensate in its equilibrium. The particle field  can be decomposed into a condensate and a noncondensate component
\be \label{ConNonCon}
  \psi({\mathbf r}, t) = \Phi({\mathbf r}, t) + {\tilde \psi}({\mathbf r}, t).
\ee
By the definition of the condensate (\ref{TDCond}), the noncondensate component  has to satisfied the condition:
\be
\langle {\tilde \psi}({\mathbf r}, t)\rangle=0.
\ee
By applying the decomposition (\ref{Decomp})  to the grand-canonical Hamiltonian, it can be separated to a quadratic and a quartic term: $K=K_2+K_4$, where
\be
  \begin{split}
   K_2 = & \int {\mbox{d}}^d{\mathbf r} \left( \tilde{\psi}^{\dagger}({\mathbf r},t) H_0 \tilde{\psi}({\mathbf r},t) \right. \\
   &  + 2 g_d \mid \Phi({\mathbf r},t)\mid^2
  \tilde{\psi}^{\dagger}({\mathbf r},t) \tilde{\psi}({\mathbf r},t)  \\
    & + \frac{g_d}{2} \Phi^2  \tilde{\psi}^{\dagger}({\mathbf r},t) \tilde{\psi}^{\dagger}({\mathbf r},t) \\ & \left.+   \frac{g_d}{2} {\Phi^{\star}}^2  \tilde{\psi}({\mathbf r},t) \tilde{\psi}({\mathbf r},t)\right) ,
  \end{split}
\ee
and
\be\label{K4}
 K_4 = \frac{g_d}{2} \int {\mbox{d}}^d{\mathbf r}   \tilde{\psi}^{\dagger}({\mathbf r},t)  \tilde{\psi}^{\dagger}({\mathbf r},t)  \tilde{\psi}({\mathbf r},t) \tilde{\psi}({\mathbf r},t).
\ee
We are interested in the regime where the condensate slightly differs from the equilibrium state, that is,
\be \label{DeltaPhi}
  \Phi({\mathbf r}, t) = \Phi_0({\mathbf r}) + \delta  \Phi({\mathbf r}, t)
\ee
with a small fluctuation $\delta  \Phi({\mathbf r}, t)$. Expanding $K_2$ up to the linear term in $ \delta  \Phi({\mathbf r}, t)$, we can rewrite it as  $K_2=K_2^{(0)}+K_2^{(1)}$, where $K_2^{(0)}$ is the zero order term of the condensate
\be
 \begin{split}
  K_2^{(0)} = &\int {\mbox{d}}^d{\mathbf r} \tilde{\psi}^{\dagger}({\mathbf r},t) \left(H_0 +
  2g_d n_0({\mathbf r}) \right) \tilde{\psi}({\mathbf r},t)  \\
    &  + \frac{g_d}{2} n_0({\mathbf r}) \left( \tilde{\psi}^{\dagger}({\mathbf r},t) \tilde{\psi}^{\dagger}({\mathbf r},t)  + \tilde{\psi}({\mathbf r},t)  \tilde{\psi}({\mathbf r},t)\right)
  \end{split}
\ee
with the condensate density $n_0({\mathbf r})=|\Phi_0({\mathbf r})|^2$, while $K_2^{(1)}$ is the linear term in the fluctuation:
\be \label{K20}
  \begin{split}
      K_2^{(1)} = &\int {\mbox{d}}^d{\mathbf r}  2g_d\Phi_0({\mathbf r}) [\delta \Phi_0({\mathbf r},t) + \delta \Phi_0^{\star}({\mathbf r},t)]  \\ & \tilde{\psi}^{\dagger}({\mathbf r},t) \tilde{\psi}({\mathbf r},t)
    + g_d \Phi_0({\mathbf r}) \left[\delta \Phi_0({\mathbf r},t) \tilde{\psi}^{\dagger}({\mathbf r},t) \tilde{\psi}^{\dagger}({\mathbf r},t)\right.\\
   & \left.+ \delta \Phi_0^{\star}({\mathbf r},t)  \tilde{\psi}({\mathbf r},t) \tilde{\psi}({\mathbf r},t) \right].
   \end{split}
\ee

As for the quartic term $K_4$, the mean-field decomposition is first used
\be \label{MF}
   \tilde{\psi}^{\dagger} \tilde{\psi}^{\dagger} \tilde{\psi} \tilde{\psi} = 4
   \tilde{n}  \tilde{\psi}^{\dagger} \tilde{\psi} +  \tilde{m} \tilde{\psi}^{\dagger} \tilde{\psi}^{\dagger}  +  \tilde{m}^{\star}  \tilde{\psi} \tilde{\psi},
\ee
where
\bea
  \tilde{n}({\mathbf r},t) & =  & \langle \tilde{\psi}^{\dagger}({\mathbf r},t) \tilde{\psi}({\mathbf r},t)\rangle \nonumber\\
  \tilde{m}({\mathbf r},t) & =  & \langle \tilde{\psi}({\mathbf r},t) \tilde{\psi}({\mathbf r},t)\rangle .
\eea
as the normal and abnormal time-dependent density.
  Under the linearization in the fluctuation (\ref{DeltaPhi}), the normal and abnormal density are also displaced with a small fluctuation  as
\bea \label{deltanm}
   \tilde{n}({\mathbf r},t) & = &\tilde{n}^{0}({\mathbf r}) + \delta\tilde{n}({\mathbf r},t), \nonumber \\
\tilde{m}({\mathbf r},t) & =  & \tilde{m}^{0}({\mathbf r}) + \delta\tilde{m}({\mathbf r},t),
\eea
expanding around their stationary values $\tilde{n}^{0}({\mathbf r}) =  \langle \tilde{\psi}^{\dagger}({\mathbf r},t) \tilde{\psi}({\mathbf r},t)\rangle_0$ and $\tilde{m}^{0}({\mathbf r}) =  \langle \tilde{\psi}({\mathbf r},t) \tilde{\psi}({\mathbf r},t)\rangle_0$. In the literature, often referred to so-called Popov approximation 
 \be \label{m0}
   \tilde{m}^{0}({\mathbf r}) =0
 \ee
 has been used in the mean-field treatment. 
Atually, this  approximation was never suggested by Popov, as indicated by Yukalov \cite{Yukalov},  but was first proposed by Shohno \cite{Shohno} and we will refer it to Shohno approximation or Shohno Ansatz in the remaining paper. 
The Shohno approximation is necessary in the  treatment  because  without that the elementary excitation would have a gap, which disobeys the gapless spectrum of the Goldstone modes caused by the continuous gauge symmetry breaking in the ground state. However, the using of Shohno approximation  is still under debate. Several attempts have been done to go beyond the Shohno approximation (for example, see Ref. \cite{Yukalov,Morgan, Watcher}).
The Popov approximation is needed in the mean-field factorization (\ref{MF}). By avoiding factorization, for example, using the perturbation or random-phase approximation to calculate the quartic terms, the gapless dispersion can be obtained even without Popov approximation.

Inserting Eqs.(\ref{DeltaPhi}), (\ref{MF}), (\ref{deltanm}) and the Shohno Ansatz (\ref{m0}) into (\ref{K4}), the quartic term $K_4$ can  be expanded up to the first order terms $K_4 = K_4^{(0)} + K_4^{(1)}$ in fluctuations $\delta{\tilde{n}}$ and  $\delta{\tilde{m}}$ as
\be
    K_4^{(0)} =  2 g_d  \int {\mbox{d}}^d{\mathbf r} \tilde{n}^0({\mathbf r})
      \tilde{\psi}^{\dagger}({\mathbf r},t) \tilde{\psi}({\mathbf r},t)
\ee
 is the zero order term, which represents the coupling to the condensate from the quartic term,  while the first order term reads
\be
  \begin{split}
  K_4^{(1)} =  & \frac{g_d}{2} \int {\mbox{d}}^d{\mathbf r} \left( 4\delta{\tilde{n}}({\mathbf r},t) \tilde{\psi}^{\dagger}({\mathbf r},t) \tilde{\psi}({\mathbf r},t) \right. \\
   + & \left. \delta{\tilde{m}}({\mathbf r},t) \tilde{\psi}^{\dagger}({\mathbf r},t) \tilde{\psi}^{\dagger}({\mathbf r},t) + \delta{\tilde{m}}^{\star}({\mathbf r},t) \tilde{\psi}({\mathbf r},t) \tilde{\psi}({\mathbf r},t)  \right).
  \end{split}
\ee
  Unlike $K_2^{(1)}$ represents the coupling between the fluctuations of the condensate and the noncondensate particles, $K_4^{(1)}$ is related to  the coupling of the noncondensate particles to the normal and abnormal densities.  If the density of the noncondensate particles is much smaller than the density of the condensate, $K_2^{(1)}$ is more important than $K_4^{(1)}$, therefore $K_4^{(1)}$ can be neglected and $K = K_2^{(0)} + K_4^{(0)} + K_2^{(1)}$.

In the case of large occupation number of particles in the condensate, $K_2^{(1)}$ is much smaller than $K_2^{(0)}$ and $K_4^{(0)}$, we can use $ K_2^{(0)} + K_4^{(0)}$ as basis to develop a perturbation expansion in terms of  $K_2^{(1)}$. To diagonalize  $K_2^{(0)} + K_4^{(0)}$, one can apply a Bogoliubov transformation
\bea \label{BogoTF}
   \tilde{\psi}({\mathbf r},t) & = & \sum_j u_j(\mathbf r) \alpha_j(t) + v_j^{\star}(\mathbf r) \alpha_j^{\dagger}(t) \nonumber \\
     \tilde{\psi}^{\dagger}({\mathbf r},t) & = & \sum_j u_j^{\star}(\mathbf r) \alpha_j^{\dagger}(t) + v_j(\mathbf r) \alpha_j(t),
\eea
 with the quasi-particle creation and annihilation operators $\alpha_j^{\dagger}$, $\alpha_j$ obeying the Bose commutation relations $[\alpha_i, \alpha_j^{\dagger}] = \delta_{ij}$, which gives the normalization condition for the functions $u_j({\mathbf r},t)$,  $v_j({\mathbf r},t)$ as
\be
   \int {\mbox{d}}^d{\mathbf r} \left[u_i^{\star}({\mathbf r})  u_j({\mathbf r}) - v_i^{\star}({\mathbf r})  v_j({\mathbf r}) \right] = \delta_{ij}.
\ee
Therefore the operator $K_2^{(0)} + K_4^{(0)}$ can be diagonalized if the Bogoliubov-de Genes equations are satisfied:
\bea \label{BogodeGen}
  {\cal L} u_j({\mathbf r})  +  g n_0({\mathbf r}) v_j ({\mathbf r}) &=& \epsilon_j u_j({\mathbf r}), \nonumber \\
   {\cal L} v_j({\mathbf r})  +  g n_0({\mathbf r}) u_j ({\mathbf r}) &=& -\epsilon_j v_j({\mathbf r}),
\eea
   where a Hermitian operator is introduced as
\be
  {\cal L} = H_0 + 2 g_d n({\mathbf r})
\ee
with the total density $n({\mathbf r})$ defined as the sum of the  condensate density and normal density in the equilibrium:  $n({\mathbf r}) = n_0({\mathbf r}) + n^0({\mathbf r})$.  As a result, the grand-canonical Hamiltonian (\ref{KOrigin}) becomes
\be \label{pertH}
  K = K_2+K_4 = \sum_j \epsilon_j \alpha_j^{\dagger} \alpha + K_2^{(1)}
\ee
with the eigenvalues $\epsilon_j$ obtained from the Bogoliubov-de Gennes equations (\ref{BogodeGen}).

In order to obtain  the decay rate,  we have  to find the time evolution of the fluctuation of the  condensate: $\delta\Phi({\mathbf r},t)$. The equation of motion:
\be
 i \frac{\partial}{\partial t} \psi({\mathbf r},t) = \left[\psi({\mathbf r},t),K\right]
 \ee
leads to the result
\be \label{EqnMot}
   i \frac{\partial}{\partial t} \psi({\mathbf r},t) = H_0 \psi({\mathbf r},t) + g_d  \psi^{\dagger}({\mathbf r},t) \psi({\mathbf r},t) \psi({\mathbf r},t).
\ee
Inserting the decomposition (\ref{ConNonCon}) into the equation of motion
  (\ref{EqnMot}), it reads
\be \label{EOM}
 \begin{split}
     i \frac{\partial}{\partial t} \Phi({\mathbf r},t) = & \left(H_0  + g_d
    |\Phi({\mathbf r},t)|^2 \right)  \Phi({\mathbf r},t) \\
      + & 2 g_d  \Phi({\mathbf r},t) \tilde{n}({\mathbf r},t) + g_d  \Phi^{\star}({\mathbf r},t) \tilde{m}({\mathbf r},t)
 \end{split}
\ee
Here we assume that the cubic product of the noncondensate contributes very little to the dynamics of the condensate, therefore the average value is set equal to zero:
$\langle \tilde{\psi}^{\dagger} \tilde{\psi} \tilde{\psi} \rangle =0$. The wavefunction in the equilibrium can be obtained by setting $\partial  \Phi_0 /\partial t= 0$, which leads to the stationary equation
\be \label{station}
  \left(H_0 + g_d\left[n_0({\mathbf r}) + 2 \tilde{n}^0({\mathbf r}) \right]\right) \Phi_0({\mathbf r}) = 0.
\ee
 By inserting Eq.(\ref{DeltaPhi}) and the stationary equation (\ref{station}) into the equation of motion for the condensate (\ref{EOM}), the equation of motion for the small amplitude $\delta\Phi$ reads
\be \label{EOMDeltaPhi}
   \begin{split}
       i \frac{\partial}{\partial t} \delta \Phi({\mathbf r},t) = & \left(H_0  + 2 g_d  n({\mathbf r})  \right)  \delta\Phi({\mathbf r},t) +
          g_d  n_0({\mathbf r}) \delta\Phi^{\star}({\mathbf r},t) \\+ & 2 g_d \Phi_0\delta\tilde{n}({\mathbf r},t) + g_d \Phi_0({\mathbf r}) \delta\tilde{m}({\mathbf r},t).
    \end{split}
\ee
 Applying the Bogoliubov transformation (\ref{BogoTF}) to Eq. (\ref{EOMDeltaPhi}), the Eq. (\ref{EOMDeltaPhi}) gives  the final form:
\be \label{EOMdeltaPhifg}
  \begin{split}
        i \frac{\partial}{\partial t} \delta \Phi({\mathbf r},t) = & \left(H_0  + 2 g_d  n({\mathbf r})  \right)  \delta\Phi({\mathbf r},t) +
          g_d  n_0({\mathbf r}) \delta\Phi^{\star}({\mathbf r},t)  \\
       & + g_d \Phi_0({\mathbf r}) \sum_{ij} \left\{ 2[u_i^{\star} u_j + v_i^{\star} v_j+ v_i^{\star} u_j] f_{ij}(t)\right. \\
       & \left. + [2 v_iu_j + u_i u_j] g_{ij}(t) +  [2 v_i^{\star}u_j^{\star} + u_i^{\star} u_j^{\star}] g_{ij}^{\star}(t) \right\},
  \end{split}
\ee
where $f_{ij}(t) \equiv \langle \alpha_i^{\dagger}(t) \alpha_j(t)\rangle$ and $g_{ij} \equiv \langle \alpha_i(t) \alpha_j(t) \rangle$ are normal and anomalous quasiparticle distribution functions.

 To calculate the normal and anomalous quasiparticle distribution functions using the perturbation  Hamiltonian (\ref{pertH}), we have to use the equation of motion :
\bea
   i \frac{\partial}{\partial t}  f_{ij}(t) & = & \langle [\alpha_i^{\dagger}(t) \alpha_j(t), K] \rangle \nonumber \\
  i \frac{\partial}{\partial t}  g_{ij}(t) & = & \langle [\alpha_i(t) \alpha_j(t), K] \rangle.  \\
\eea
To the first order, the Fourier transform of $f_{ij}$ and $g_{ij}$ at the frequency $\omega$ is given by
\be \label{fij}
  \begin{split}
    f_{ij}(\omega) = & 2 g_d  \,\frac{f_i^0 - f_j^0}{\omega + (\epsilon_i - \epsilon_j) + i 0^{+}} \int {\mbox{d}}^d{\mathbf r} \\
      & \times \Phi_0 \left[\delta\Phi_1({\mathbf r},\omega) (u_iu_j^{\star} + v_iv_j^{\star} + v_i u_j^{\star}) \right.\\
      & +  \left.\delta\Phi_2({\mathbf r},\omega) (u_iu_j^{\star} + v_iv_j^{\star} + u_i v_j^{\star}) \right];
  \end{split}
\ee
\be\label{gij}
  \begin{split}
    g_{ij}(\omega) = & 2 g_d  \,\frac{1+f_i^0 + f_j^0}{\omega - (\epsilon_i + \epsilon_j) + i 0^{+} } \int {\mbox{d}}^d{\mathbf r} \\
      & \times \Phi_0 \left[\delta\Phi_1({\mathbf r},\omega) (u_i^{\star}v_j^{\star} + v_i^{\star} u_j^{\star} + u_i^{\star} u_j^{\star}) \right.\\
      & +  \left.\delta\Phi_2({\mathbf r},\omega) (u_i^{\star} v_j^{\star} + v_i^{\star} u_j^{\star} + v_i^{\star} v_j^{\star}) \right],
  \end{split}
\ee
where $\delta\Phi_1({\mathbf r},\omega)$ and  $\delta\Phi_2({\mathbf r},\omega)$are the Fourier transform of $\delta\Phi({\mathbf r},t)$ and $\delta\Phi^{\star}({\mathbf r},t)$:
\bea
  \delta\Phi_1({\mathbf r},\omega) & = & \int dt e^{-i\omega t} \delta\Phi({\mathbf r},t) \nonumber \\
  \delta\Phi_2({\mathbf r},\omega) & = & \int dt e^{-i\omega t} \delta\Phi^{\star}({\mathbf r},t),  \label{FTPhi} \\
\eea
and $f_j^0$ is the bosonic distribution function
\be
  f_j^0 = \frac{1}{[e^{\beta\epsilon_j}-1]}
\ee
with $\beta = 1/k_BT$.  Fourier transforming the equation of motion
 (\ref{EOMdeltaPhifg}) and replacing Eqs. (\ref{fij}) and (\ref{gij}) into it, we obtain the perturbed eigenfrequency:
\be \label{omega}
  \begin{split}
  \omega = & \omega_0 + 4 g_d^2 \sum_{ij} (f_i^0-f_j^0) \, \frac{|A_{ij}|^2}{\omega_0 + (\epsilon_i-\epsilon_j) + i0^{+}} \\
           & + 2 g_d^2  \sum_{ij} \left(1+ f_i^0 + f_j^0) \, \frac{|B_{ij}|^2}{\omega_0 - (\epsilon_i + \epsilon_j) + i0^{+}}\right. \\
   & \left. - \frac{|\tilde{B}_{ij}|^2}{\omega_0+(\epsilon_i+\epsilon_j) + i0^{+}}\right),
  \end{split}
 \ee
 where the unpeturbed eigenfrequency $\omega_0$ is obtained from the unperturbed RPA equation \cite{Hutchinson}
 \be \label{Bogofluctuations}
   \left( \begin{array}{cc} {\cal L} & g_d n_0 \\ -g_d n_0 & -{\cal L} \end{array}\right) \left(\begin{array}{c} \delta\Phi_1^0 \\ \delta \Phi_2^0 \end{array}\right) = \omega_0  \left(\begin{array}{c} \delta\Phi_1^0 \\ \delta \Phi_2^0 \end{array}\right)
 \ee
with the normalization condition
\be
  \int {\mbox{d}}^d{\mathbf r} (|\delta\Phi_1^0|^2 -|\delta\Phi_2^0|^2 ) = 1,
\ee
and $A_{ij}$, $B_{ij}$ and $\tilde{B}_{ij}$ are defined as
\be \label{AB}
 \begin{split}
    A_{ij} = & \int {\mbox{d}}^d{\mathbf r}
      \Phi_0 \left[\delta\Phi_1^0 (u_iu_j^{\star} + v_iv_j^{\star} + v_i u_j^{\star}) \right.\\
      & +  \left.\delta\Phi_2^0 (u_iu_j^{\star} + v_iv_j^{\star} + u_i v_j^{\star}) \right], \\
     B_{ij} = & \int {\mbox{d}}^d{\mathbf r} \Phi_0 \left[\delta\Phi_1^0 (u_i^{\star}v_j^{\star} + v_i^{\star} u_j^{\star} + u_i^{\star} u_j^{\star}) \right.\\
      & +  \left.\delta\Phi_2^0  (u_i^{\star} v_j^{\star} + v_i^{\star} u_j^{\star} + v_i^{\star} v_j^{\star}) \right] \\
 \tilde{B}_{ij} = & \int {\mbox{d}}^d{\mathbf r} \Phi_0 \left[\delta\Phi_1^0(u_iv_j+ v_i u_j + u_i u_j) \right.\\
      & +  \left.\delta\Phi_2^0 (u_i v_j + v_i u_j + v_i v_j) \right]. \\
  \end{split}
\ee
The real part of the right-hand side (Eq. (\ref{omega}))  gives the eigenenergy of the system and the imaginary part tells us the damping coefficient $\gamma$. Using the relation
\be
 \frac{1}{x+i0^{+}} = P\frac{1}{x} - i\pi \delta(x),
\ee
we can divide the damping rate into two different types: one comes from the process that one phonon with the frequency $\omega_0$ is absorbed by a thermal excitation $\epsilon_i$ jumping to another thermal excitation with the energy $\epsilon_j = \epsilon_i + \omega_0$. This mechanism  is so-called Landau damping given by the second term on the right-hand side of Eq. (\ref{omega}),
\be \label{LD}
  \gamma_L = 4 \pi g_d^2 \sum_{ij} |A_{ij}|^2 (f_i^0-f_j^0) \delta(\omega_0 + \epsilon_i - \epsilon_j).
\ee
This process happens mostly at finite temperature,  it is therefore a thermal process.
Another kind of decay arises from the process of a long wave-length phonon decaying into two phonons, as indicated by Beliaev, and it can be obtained by the imaginary part of the first term in brackets on the right-hand side of Eq. (\ref{omega}):
\be\label{BD}
  \gamma_B = 2 \pi g_d^2 \sum_{ij} |B_{ij}|^2 (1+f_i^0+f_j^0) \delta(\omega_0 - \epsilon_i - \epsilon_j).
\ee
This process occurs mostly at zero temperature, which is a pure quantum effect.
The total damping rate is the sum of the two damping coefficients: $\gamma = \gamma_B+\gamma_L$.

In this paper we are interested in homogeneous Bose gases, i.e. $V_{\mbox{ext}}({\mathbf r}) = 0$. For homogeneous systems the condensate density remains the same throughout the space: $\Phi_0 = \sqrt{n_0}$, while the excitations and the fluctuations satisfying Eq. (\ref{BogodeGen}) and Eq. (\ref{Bogofluctuations}) can be described by the plane waves
\be
  \left( \begin{array}{c} \delta\Phi_1({\mathbf r}) \\ \delta\Phi_2({\mathbf r})  \end{array}\right) = \frac{1}{\sqrt{V}} \int {\mbox{d}}^d {\mathbf k} e^{i{\mathbf{k}}\cdot{\mathbf{r}}}\left( \begin{array}{c} u_{k} \\ v_{k} \end{array}\right),
\ee
 \be
  \left( \begin{array}{c}  u_{\mathbf{k}'}({\mathbf r}) \\ v_{\mathbf{k}'}({\mathbf r})  \end{array}\right) = \frac{1}{\sqrt{V}} \int {\mbox{d}}^d {\mathbf k'}e^{i{\mathbf{k}'}\cdot{\mathbf{r}}} \left( \begin{array}{c} u_{{k}'} \\ v_{{k}'} \end{array}\right),
\ee
where $u_{k}$ and  $v_{k}$ satisfy the Bogoliubov relations:
\be
  \begin{split}
  u_{k}^2 = 1+ v_{k}^2 = & \frac{(\epsilon^2(\mathbf{k})+g_d^2 n_0^2)^{1/2}+\epsilon(\mathbf{k})}{2\epsilon(\mathbf{k})} \\
   u_{k} v_{k} = & -\frac{g_d n_0}{2 \epsilon(\mathbf{k})}
  \end{split}
\ee
and $\epsilon(\mathbf{k})$ is the Bogoliubov energy (\ref{BogoliubovSpectrum}).

\subsection{Quantum regime}

As mentioned in the last section, the dacay rate is mostly contributed by Beliaev damping, which can be obtained by setting $f_i^0=f_j^0 = 0$ in Eq. (\ref{BD}). The matrix element $B_{{\mathbf k}' - {\mathbf k}, {\mathbf k}'}$ reads
\be \label{Bmatrix}
   \begin{split}
     B_{{\mathbf k}' - {\mathbf k}, {\mathbf k}'} = & \sqrt{\frac{n_0}{V}} \left[(u_{k} (u_{k'}v_{k'-k}+v_{k'}u_{k'-k} + u_{k'} u_{k'-k})  \right. \\
     & \left.+ v_k(u_{k'} v_{k'-k} + v_{k'} u_{k'-k} + v_{k'} v_{k'-k})\right],
   \end{split}
\ee
and other elements are zero.
At zero temperature, only the momenta  with long wavelength are involved in the Beliaev damping process, i.e. $k'\sim k \sim |{\mathbf k}'- {\mathbf k}| \ll mc$. Therefore the long-wavelength approximation for the energy $\epsilon(\mathbf{k})$  and the wave functions $u_k$ and $v_k$ can be used:
\be \label{epzeroT}
  \epsilon(\mathbf{k}) \simeq ck + \frac{k^3}{8m^2c},
\ee
\bea\label{uvzeroT}
  u_k & \simeq & \left(\frac{mc}{2k}\right)^{1/2} + \frac{1}{2}\left(\frac{k}{2mc}\right)^{1/2}+ \frac{1}{8}\left(\frac{k}{2mc}\right)^{3/2} \nonumber \\
& &  -\frac{1}{8}\left(\frac{k}{2mc}\right)^{5/2} \nonumber \\
  v_k & \simeq & -\left(\frac{mc}{2k}\right)^{1/2} + \frac{1}{2}\left(\frac{k}{2mc}\right)^{1/2} - \frac{1}{8}\left(\frac{k}{2mc}\right)^{3/2} \nonumber \\ & & -\frac{1}{8}\left(\frac{k}{2mc}\right)^{5/2}.
\eea
Substituting (\ref{epzeroT}) and (\ref{uvzeroT}) in Eq. (\ref{Bmatrix}), we obain the result
\be \label{BME}
   B_{{\mathbf k}' - {\mathbf k}, {\mathbf k}'} =  \sqrt{\frac{n_0}{V}}
   \frac{3}{4\sqrt{2}} \frac{|{\mathbf k}| |{\mathbf k}'||{\mathbf k}'-{\mathbf k}|}{(mc)^{3/2}}.
\ee
Inserting Eq. (\ref{BME}) to  Eq. (\ref{BD}) and summerizing all the momenta ${\mathbf k}$ and  ${\mathbf k}'$,   one obtain the same result (\ref{eq:BD}) as that  using hydrodynamical approach. Therefore we reproduces the results for $3$-D and $2$-D decay as Eq. (\ref{3DBD}) and Eq. (\ref{2DBD}).

\subsection{Thermal Regime}
In the theraml regime where $\omega_0 \simeq k_B T$, a long-wavelength Goldstone mode with the eigenfrequency ($\omega_0 \simeq ck$) describes the behavior of the condensate in the thermal clouds. In this limit,  the $u$ and $v$ functions can be expanded as
\bea
  u_k  & \simeq & \left( \frac{mc^2}{2 \epsilon(k)}\right)^{1/2} + \frac{1}{2} \left( \frac{\epsilon(k)}{2mc^2}\right)^{1/2},  \nonumber \\
  v_k  & \simeq & -\left( \frac{mc^2}{2 \epsilon(k)}\right)^{1/2} + \frac{1}{2} \left( \frac{\epsilon(k)}{2mc^2}\right)^{1/2}.
\eea
Using this expansion,the long-wavelength behavior for the nonzero elements of the matrix $A$  can be expressed  as
\be \label{AmatrixLD}
  \begin{split}
     A_{{\mathbf k}',{\mathbf k}'+{\mathbf k}} = & \frac{\sqrt{n_0}}{\sqrt{V}}
     \left(\frac{\epsilon(k)}{2mc^2}\right) ^{1/2} \left(u_{k'}^2 + v_{k'}^2+ u_{k'} v_{k'} \right. \\
 & \left. -\frac{v_g}{c} \cos{\theta} \frac{2 u_{k'}^2 v_{k'}^2}{u_{k'}^2 + v_{k'}^2} \right),
  \end{split}
\ee
with the angle $\theta$ between the vectors ${\mathbf k}'$ and ${\mathbf k}$, and the group velocity of the excitation defined as $v_g = \partial \epsilon_{k}/\partial k$.

In three dimensions, the damping rate can be obtained by inserting the nonzero coefficients (\ref{AmatrixLD})  into (\ref{LD}) and integrating out the angle $\theta$ as follows:
\be
  \frac{\gamma^{d=3}}{\epsilon(k)} \simeq \frac{\gamma_L^{d=3}}{\epsilon(k)} = (a_3^3 n_0)^{1/2} F(\tau)
\ee
 where $\tau = k_B T/mc^2$ as dimensionless temperature, $a_3$ is the three-dimensional scattering length, and $F(\tau)$ is defined in the following:
\be
  F(\tau) = \frac{\sqrt{\pi}}{\tau} \int dz \; \sech^{2}(\frac{z}{2\tau})  (1-\frac{1}{2u}-\frac{1}{2u^2})^2
\ee
with the definition $u=\sqrt{1+z^2}$. This expression was first found by Pitaevskii and Stringari \cite{Pitaevskii}.

For low twmperature $ k_BT \ll mc^2$, i.e. $\tau \ll 1$, the function $F$ takes its limit $F \approx 3 \pi^{9/2} \tau^4 /5$ and one finds the Hohenberg and Martin's result (\ref{Hohenberg}).  As mentioned in the last section, the hydrodynamical approach and HFB give the same limit at low temperature. However, at high temperature, the hydrodynamic approach fails. For temperature $\tau \gg 1$., i.e. $k_BT \gg mc^2$, the function $F$ takes the asymptotic limit $F \rightarrow 3 \pi^{3/2} \tau /4$, and the damping rate approaches the  Sz{\'e}pfalusy and Kondor result (\ref{Kondor}).  Therefore the HFB gives correct asymptotic value at high temperature.

In Fig. \ref{fig3} the famous result for the three-dimensional damping rate using HFB method obtained by Stringari and Pitaevskii \cite{Pitaevskii}, and then recovered by Giorgini \cite{Giorgini} is shown as the solid line. We can see that the three-dimensional damping rate leaves the low-temperature limit very soon, while it approaches the high-temperature linear law very slowly.

In two dimensions, HFB offers a good approximation for all regime of temperature. After inserting the matrix elements (\ref{AmatrixLD})  into {\ref{LD}}  and then integrating out the angle $\theta$, one obtains the damping rate:
\be
   \frac{\gamma^{d=2}}{\epsilon(k)} \simeq \frac{\gamma_L^{d=2}}{\epsilon(k)} = 2 mg_2 G(\tau),
\ee
where
\be
 \begin{split}
      G(\tau) = &  \frac{\sqrt{2}}{\pi\tau} \int_0^{\infty}
            dz  \frac{\sech^{2}(\frac{z}{2\tau})}{\sqrt{\sqrt{z^2+1}-1-\frac{z^2}{2(z^2-1)}}} \\
     &  \left(1-\frac{1}{2u}-\frac{1}{2u^2}\right)^2.
   \end{split}
\ee
In the low temperature limit $\tau \ll 1$, the function $G$ takes the limit: $G(\tau) \rightarrow \sqrt{3} \pi \tau^2 /16$,  the damping rate goes to the result (\ref{2DLDLT}). As in three dimensions HFB  gives the same result as the hydrodynamical approach at low temperature in two dimensions.

For high temperature $k_BT \gg mc^2$, the function $G$ takes its asymptotic value: $G(\tau) \rightarrow 0.013 \tau$, and the damping rate is given by
\be
  \frac{\gamma^{d=2}}{\epsilon(k)} \simeq 0.026 \frac{m k_B T}{n_0}.
\ee
Therefore the damping rate itself  reads
\be
    \gamma^{d=2} \simeq 0.013 \frac{8\pi}{|\ln{(1/n a_2^2)}|} \frac{k k_B T}{m c}.
\ee
In Fig. \ref{fig4}, the two-dimensional damping rate per unit energy using HFB method is also shown as a function of $\tau$ (solid line). We can see that the two-dimensional damping rate approaches the high temperature linear law much sooner than the  three-dimensional case. That means, the two dimensional systems go to the high-temperature  asymptotic value at lower value of temperature compared to that in three dimensions. This behavior has been found by Guilleumas and Pitaevskii studying a quasi two-dimensional system \cite{2DPi} . In this figure one can also see that the hydrodynamical approach can only give good results in the regime where $\tau < 0.5$.

\section{Conclusion}
In this work, we have compared the hydrodynamical approach and the HFB approach to calculate the damping rate in 2D and 3D bose gas. The hydrodynamical approach is a powerful tool due to the fact that one can use the  Green's-function technique based on the effective action (\ref{eq:Seff}). This works very well at zero and low temperatures. However, this method truncates the rapid oscillating fields, which is not the case for the high temperature regime,therefore it overestimates the Landau damping. On the other hand, the HFB approximation can explain either low temperature or high temperature damping. It seems that the mean-field approach (HFB) is a better method for Bose gases. The HFB approach based on the mean field method factorizes the quartic terms and therefore Shohno Ansatz has to be used to avoid anomalous behavior. In the absence of the Shohno Ansatz, there would exist a gap in the low excitation spectrum. Therefore the mean field approach cannot guarantee a zero energy gap. From a physical point of view, the existence of a gapless excitation is a general rule for  Bose systems. In order to avoid errors in the higher order calculations, the mean field approach should be very carefully used. Therefore we can see the benefit of the hydrodynamical approach for the low temperature regime. The low energy excitation is always gapless using hydrodynamical approach.  Another benefit of using the hydrodynamical approach, as indicated by Popov \cite{Popov}, is that it can avoid the strange divergence at high and low momenta, so-called ultraviolet and infrared catastrophe, which can be caused by the perturbation theory based on the mean-field approach.

We have found for the first time that the Beliaev damping rate is proportional to $k^3$ at zero temperature and the Landau damping rate for the 2D bose gas is proportional to $ T^2$ for low temperature and to $T$ for high temperature. The behavior of the 2D damping is also totally different from the 3D damping. While the 3D Landau damping approaches the linear regime very slowly with increasing temperature, the 2D damping become linear very fast. The linear regime symbolizes the classical high temperature behavior, therefore the two dimensional systems go to the high-temperature  asymptotic value at lower value of temperature compared to that in three dimensional system. This behavior was also found in Ref. \cite{2DPi} with numerical calculation for a quasi-2D system.

\acknowledgements 
 We especially thank A. G. Semenov for the correction of the two-dimensional damping rate, and  V. I. Yokalov for the kindly reminding of the unjust using of the concept ``Popov approximation''. 
Ming-Chiang Chung is supported by the National Science Coucil of Taiwan, R.O.C. under grant No:NSC95-2112-M001-054-MY3. 


\end{document}